\documentclass[
    ,final            
    ,draft            
    ,4apaper
  ]
  {aipproc}

\layoutstyle{6x9}

\usepackage{graphicx}

\begin{document}


\title[Hyperfine induced transition rates in Be-like ions]{Storage-Ring Measurements of Hyperfine Induced Transition Rates in Berylliumlike Ions}

\author{Stefan Schippers}{
  address={Institute for Atomic and Molecular Physics, Justus-Liebig-University Giessen,\\ Leihgesterner Weg 217, 35392 Giessen, Germany},
  email={Stefan.Schippers@physik.uni-giessen.de}
}

\begin{abstract}
The status of experimental measurements and theoretical calculations of the hyperfine induced $2s\,2p\;^3P_0\to 2s^2\;^1S_0$ transition rate in Be-like ions is reviewed. Possible reasons, such as external electromagnetic fields and competing E1M1 two-photon transitions, for presently existing significant discrepancies between experiment and theory are discussed. Finally, directions for future research are outlined.
\end{abstract}

\maketitle

\section{Introduction}

The influence of the hyperfine interaction on atomic lifetimes was already noted as early as 1930 by Bowen who commented on the spectroscopic work of Huff and Houston \cite{Huff1930} on \lq\lq forbidden\rq\rq\ lines in mercury.
The hyperfine interaction leads to a shortening of lifetimes of metastable levels and, correspondingly, to an enhanced associated emission of radiation. In astrophysics, this can be used for discriminating between radiation from different isotopes and, thus, for inferring isotope-dependent abundance ratios which provide insight into stellar nucleosynthesis \cite{Brage1998a,Rubin2004a}.

\begin{figure}[b]
\centering{\includegraphics[width=0.8\textwidth]{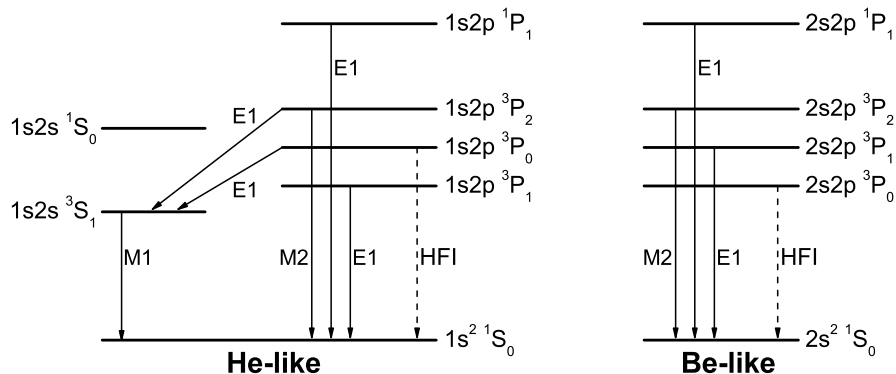}}
\caption{\label{fig:HeBeLevels}Schematic of the lowest energy levels and their main one-photon decay modes in He-like and Be-like ions. The one-photon transitions are labeled E1 (electric dipole), M2 (magnetic quadrupole), and HFI (hyperfine induced). The energies are not to scale. In the absence of a nuclear spin the hyperfine-induced (HFI) one-photon transitions are forbidden.}
\end{figure}

\begin{figure}[t]
\includegraphics[width=0.7\textwidth]{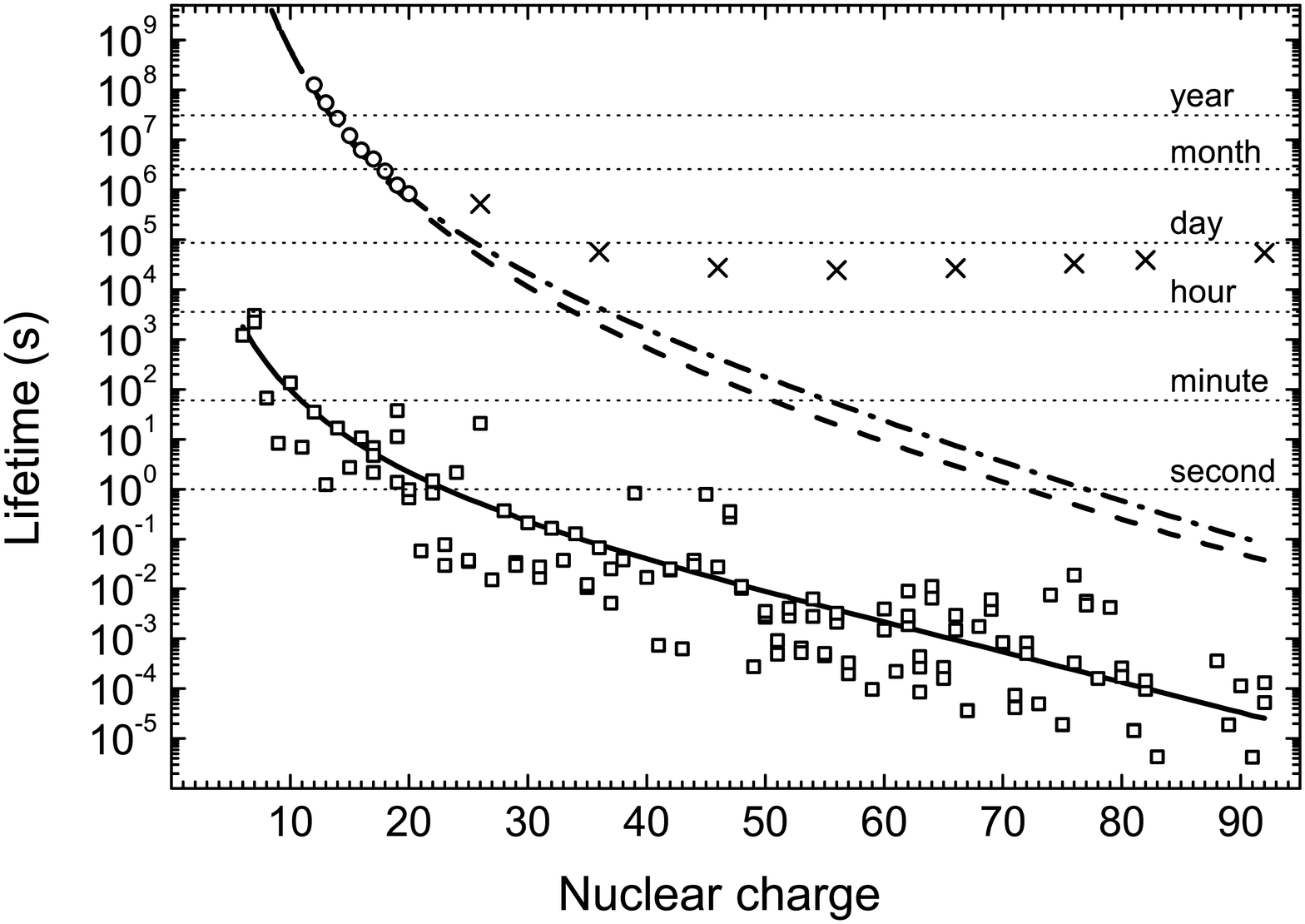}
\caption{\label{fig:thfi}Theoretical results for lifetimes associated with one-photon and two-photon $2s\,2p\;^3P_0\to 2s^2\;^1S_0$ transitions in Be-like ions. Open squares: Hyperfine-induced lifetimes by \citet{Cheng2008a}.
Full line:  Reduced HFI lifetime $\tau_\mathrm{el} = 1/A_\mathrm{el}$ calculated from the results of \citet{Cheng2008a} by using Eq.~\ref{eq:AHFIred}. Open circles: Lifetimes $\tau_{E1M1}$ associated with the
E1M1 two-photon transition calculated by \citet{Schmieder1973a}. Dashed and dash-dotted lines: $\tau_{E1M1}$ from the formula of \citet{Laughlin1980a} (Eq.~\ref{eq:AE1M1E})  and  from Eq.~\ref{eq:AE1M1D} with $E$ and $\Delta$ taken from \citep{Cheng2008a}. Crosses: Lifetimes associated with Stark quenching of the $2s\,2p\;^3P_0$ level in an electric field of $4.5\times10^8$~V/m as predicted by \citet{Maul1998a}.}
\end{figure}

Experimentally, the effect has predominantly been studied by beam-foil spectroscopy of He-like ions (for a recent comprehensive review see \cite{Johnson2011}) where the hyperfine-induced (HFI) transition of the metastable $1s\,2s\;^3P_0$ level to the $1s^2\;^1S_0$ ground level competes with the much faster transition to the $1s\,2s\;^3S_1$ first excited level (Fig.~\ref{fig:HeBeLevels}). The situation is much more clear-cut for Be-like ions, and, in general, divalent atoms and ions, having a $ns^2\;^1S_0$ ground level and a valence shell $n$,  where the $ns\,np\;^3P_0$ level is the first excited level which in isotopes with nuclear spin $I=0$ cannot decay by a one-photon transition. For $I\ne0$ the hyperfine interaction mixes levels with different $J$ and the $ns\,np\;^3P_0$ level acquires a finite radiative lifetime ($\tau_\mathrm{HFI}$), which depends strongly on nuclear charge and nuclear magnetic moment.  Theoretical predictions \citep[see Fig.~\ref{fig:thfi}]{Cheng2008a} for  Be-like ions range from $\tau_\mathrm{HFI}\approx 3000$~s for $^{15}$N$^{3+}$ down to  $\tau_\mathrm{HFI}\approx 4$~$\mu$s for $^{231}$Pa$^{87+}$.

Such long lifetimes are attractive in view of obtaining ultraprecise optical frequency standards \cite{Ludlow2008,Rosenband2008}. This has motivated experiments with trapped  In$^+$ \cite{Becker2001a} and Al$^+$  \cite{Rosenband2007a} ions which provided experimental values for $\tau_\mathrm {HFI}$ (with uncertainties of  $\sim4\%$ and $\sim7\%$, respectively) from optical spectroscopy of the atomic-clock transitions $5s\,5p\;^3P_0 \to 5s^2\;^1S_0$ in $^{115}$In$^+$ and $3s\,3p\;^3P_0 \to 3s^2\;^1S_0$ in $^{27}$Al$^+$. For Be-like $^{14}$N$^{3+}$ $\tau_\mathrm{HFI}$ was inferred from a spectroscopic observation of a planetary nebula \cite{Brage2002a}.

Another experimental environment that permits access to long atomic lifetimes is a heavy-ion storage ring \cite{Traebert2010}.  In fact, the only laboratory values for HFI lifetimes in Be-like ions that are available to date were obtained from storage ring experiments with $^{47}$Ti$^{18+}$ and $^{33}$S$^{12+}$ ions \cite{Schippers2007a,Schippers2012}. The purpose of the present paper is to summarize these investigations and to point out directions for future research.

\section{Storage-ring experiments}

\begin{figure}[t]
\includegraphics[width=0.7\textwidth]{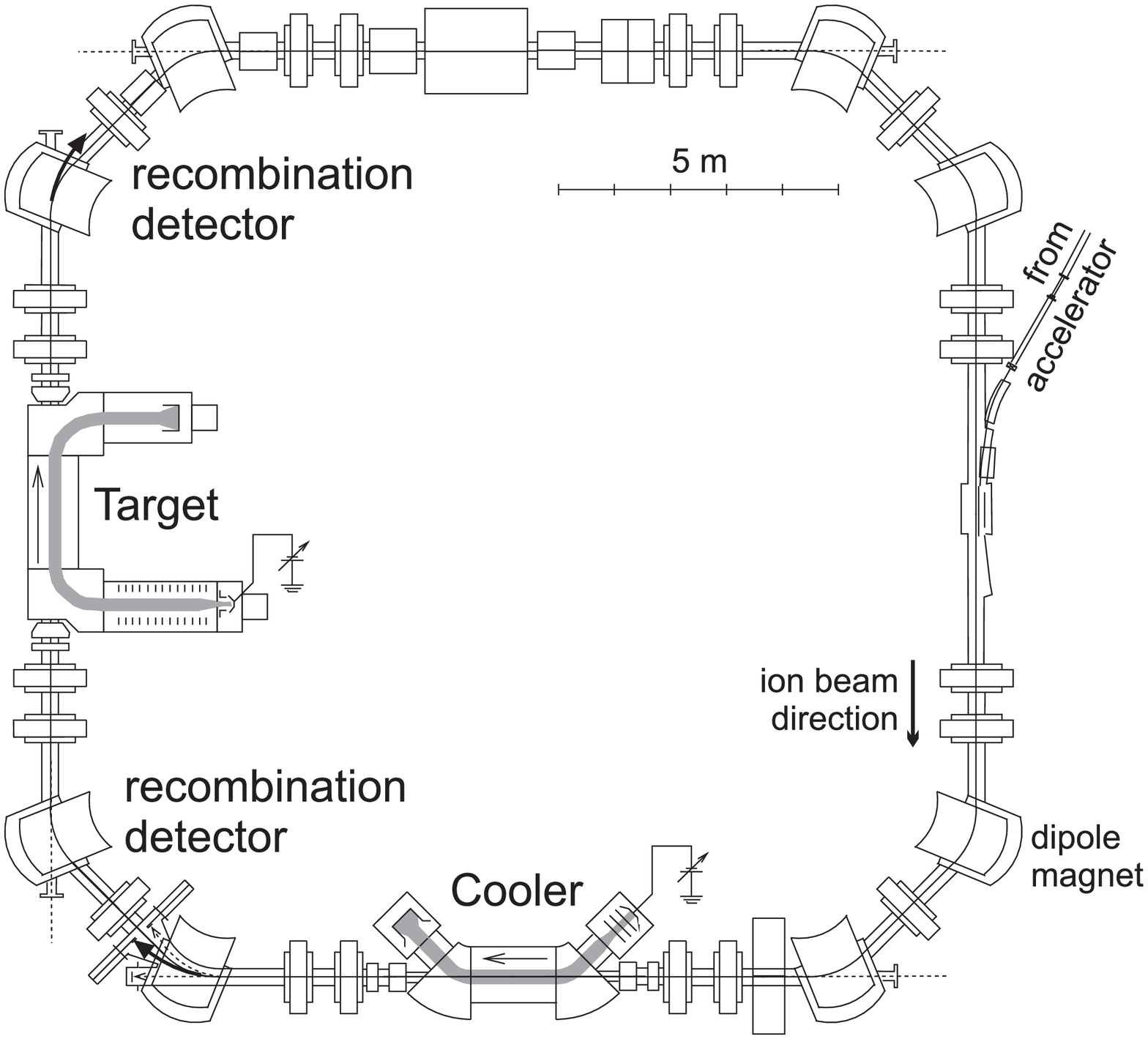}
    \caption{\label{fig:TSR}Layout of the heavy-ion storage ring TSR \citep{Grieser2012} at the Max-Planck-Institute of Nuclear Physics in Heidelberg, Germany. The ring has a circumference of 55.4~m.}
\end{figure}

Heavy-ion storage-rings offer excellent conditions for experiments with highly charged atomic ions \citep{Mueller1997c,Schuch2007a,Schwalm2007,Schippers2009}. The salient feature of a storage ring is that ions with \emph{well defined mass and charge state} can be stored for rather long times. Usually the storage time is limited by ion losses from the ring due to charge changing collisions with residual gas particles. Therefore, ultrahigh vacuum conditions are mandatory as well as high velocities --- typically $0.05c-0.3c$ (with $c$ denoting the speed of light) --- of the stored ions since the relevant atomic cross sections strongly decrease with increasing collision energy. Depending  on ion energy and charge state storage times ranging from seconds to several hours have been realized \cite{Grieser2012}.

In principle, any atomic lifetime which is significantly shorter than the storage time can be experimentally investigated. Ions in excited atomic states are usually generated by the charge stripping process that is used for producing the desired ion charge state in the accelerator prior to injection of the ions into the storage ring. Most of the excited states decay rapidly to the ground state and to long-lived excited states. The standard technique for measuring atomic decay rates in an ion storage-ring is to monitor the fluorescence from the long-lived excited states as function of storage time \cite{Traebert2010}. However, this approach suffers from small solid angles and background photons which severely hampers the investigation of weak decay channels \cite{Traebert2011}.

\begin{figure}[t]
\includegraphics[width=0.8\columnwidth]{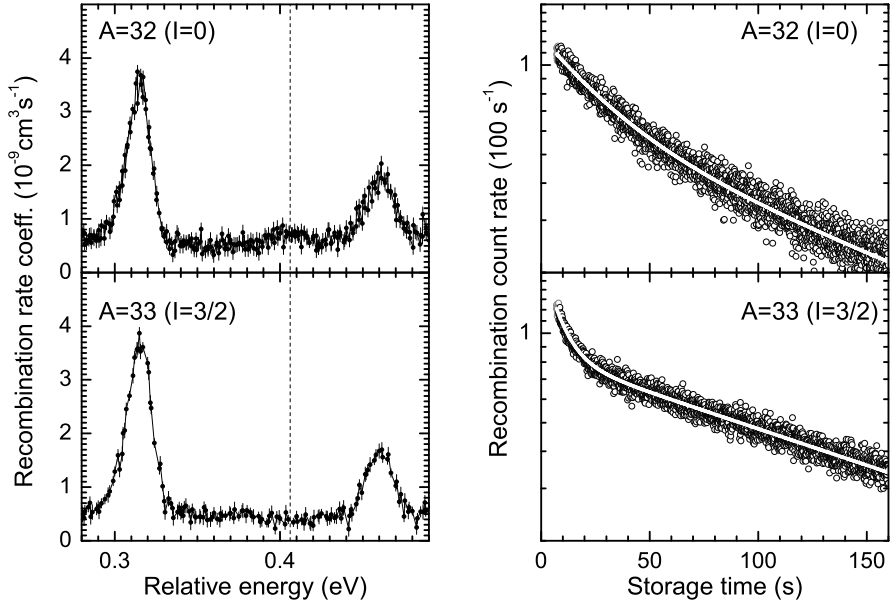}
\caption{\label{fig:S12lifetime}Left: Measured recombination spectra \citep{Schippers2012}
for $^{32}$S$^{12+}$ and $^{33}$S$^{12+}$ ions. A dielectronic recombination resonance at $\sim$0.4~eV (vertical dashed line) that is excited from
the metastable $1s^2\,2s\,2p\;^3P_0$ level  is not visible in the $^{33}$S$^{12+}$ spectrum, since
the $^3P_0$ state is quenched by the hyperfine interaction after sufficiently long storage times.
Only resonances that are excited from the $1s^2\,2s^2\;^1S_0$ ground level occur in both spectra. Right:
Measured (symbols) and fitted (lines) decay curves \citep{Schippers2012} for $^{32}$S$^{12+}$
and $^{33}$S$^{12+}$ ions stored in the TSR heavy-ion storage ring.
The monitored signal is the S$^{11+}$ production rate at an electron-ion collision energy of 0.406~eV
(marked by the arrow in the left panel). In contrast to the $A=32$ (upper) curve the $A=33$ (lower)
curve exhibits a fast decaying component which is the signature of the hyperfine induced $2s\,2p\;^3P_0 \to 2s^2\;^1S_0$ transition. A similar figure for Ti$^{18+}$ can be found in \cite{Schippers2009}.}
\end{figure}

For the measurements of the HFI $2s\,2p\; ^3P_0 \to 2s^2\; ^1S_0$ transition rates in Be-like ions an alternative approach \citep{Schmidt1994} has been applied \citep{Schippers2007a,Schippers2012} at the heavy-ion storage-ring TSR (Fig.~\ref{fig:TSR}) of the Heidelberg Max-Planck-Institute for Nuclear Physics (MPIK). In these experiments, Be-like A$^{(Z-4)+}$ ions with nuclear charge $Z$ from the MPIK accelerator facility were injected into the storage ring. The decrease of the number of stored ions was monitored by counting A$^{(Z-5)+}$ ions which were produced from the primary beam by electron-ion recombination in an electron beam (either \lq Cooler\rq or \lq Target\rq in Fig.~\ref{fig:TSR}) that moved collinearly with the ion beam in one of the straight sections of TSR. Separation of A$^{(Z-5)+}$ product ions from the A$^{(Z-4)+}$ primary ions took place in the first TSR dipole bending magnet behind the electron target. A single-particle detector (with negligible dark count-rate) was positioned at a suitable location behind that magnet such that it intercepted the path of the A$^{(Z-5)+}$ product ions. Since the ions moved at $\sim$10\% of the speed of light all reaction products were emitted into a narrow cone in the laboratory system and could be detected with essentially 100\% efficiency. This is a decisive advantage of the recombination method over the more conventional detection of fluorescence photons.

In order to selectively increase the yield from the fraction of stored ions that were in the excited $2s\,2p\;^3P_0$ level, the electron-ion collision energy was tuned to the energy  of a recombination resonance associated with this level. The left panels in figure \ref{fig:S12lifetime} show electron-ion recombination spectra of two isotopes of the same ion, i.e., $^{32}$S$^{12+}$ (upper left) and $^{33}$S$^{12+}$ (lower left) ions. The spectra differ significantly in the strength of the recombination resonance at $\sim$0.4e~eV collision energy. It is visible only in the $^{32}$S$^{12+}$ spectrum. The isotope $^{32}$S has zero nuclear spin and, correspondingly, the $^{32}$S$^{12+}$($2s\,2p\;^3P_0$) level has a practically infinite lifetime. In contrast, the isotope $^{33}$S has nuclear spin $I=3/2$ and the $^{33}$S$^{12+}$($2s\,2p\;^3P_0$) level is quenched by the hyperfine interaction such that any recombination resonance associated with this level is rapidly diminished. The right panels in figure \ref{fig:S12lifetime} show the decrease of the recombination signal --- with the electron-ion collision energy kept at 0.4~eV --- as function of storage time.

\begin{table}[t]
\caption{\label{tab:comp}Comparison of storage-ring results for the hyperfine induced $2s\,2p\; ^3P_0 \to 2s^2\; ^1S_0$ transition rate in Be-like ions with the available theoretical values. For each ion the nuclear charge $Z$, the nuclear spin $I$, and the nuclear magnetic moment $\mu$ (in units of the nuclear magneton) \citep{Stone2005a} are also given.}
\begin{tabular}{lccccc}
\hline
   \tablehead{1}{c}{b}{Ion}
 & \tablehead{1}{c}{b}{Year}
 & \tablehead{2}{c}{b}{$\mathbf{\tau_\mathrm{HFI}}$ (s)}
 & \tablehead{1}{c}{b}{Deviation from}
 & \tablehead{1}{c}{b}{Reference} \\[-1.5ex]
 &
 & \tablehead{1}{c}{b}{experiment}
 & \tablehead{1}{c}{b}{theory}
 & \tablehead{1}{c}{b}{experiment}
 & \\
     \hline\rule[0mm]{0mm}{4mm}%
 $^{47}$Ti$^{18+}$             & 1993 &                   & \phantom{2}2.812  & \phantom{-}36\% & \citet{Marques1993} \\
    ~~$Z=22$                   & 1998 &                   & \phantom{2}1.041\tablenote{calculated from Eq.~\ref{eq:AHFIred} with an interpolated value for $A_\mathrm{el}$}  & -73\%                & \citet{Brage1998a}\\
    ~~$I=5/2$                  & 2007 & \phantom{1}1.8(1) &                   &                 & \citet{Schippers2007a}\\
   ~~$\mu=-0.78848$            & 2008 &                   & \phantom{2}1.487  &           -21\% & \citet{Cheng2008a}\\
                               & 2009 &                   & \phantom{2}1.476  &           -22\% & \citet{Andersson2009}\\
                               & 2010 &                   & \phantom{2}1.513  &           -19\% & \citet{Li2010}\\
          & & & & & \\
 $^{33}$S$^{12+}$              & 1993 &                   & 27.69\phantom{6}  & \phantom{-}62\% & \citet{Marques1993} \\
    ~~$Z=16$                   & 1998 &                   & \phantom{2}8.73$^*$  &        -19\% & \citet{Brage1998a}\\
  ~~$I=3/2$                    & 2008 &                   & 10.74\phantom{6}  & \phantom{-6}3\% & \citet{Cheng2008a}   \\
 ~~$\mu=+0.64382$              & 2009 &                   & 10.69\phantom{6}  & \phantom{-6}3\% & \citet{Andersson2009}\\
                               & 2012 &          10.4(4)  &                   &                 & \citet{Schippers2012}\\
 \hline
\end{tabular}
\end{table}

For the further analysis the double exponential function
\begin{equation}\label{eq:Rate2}
 N(t) = c_m e^{\displaystyle  -t/\tau_m}+ c_g e^{\displaystyle -t/\tau_g}
\end{equation}
was fitted to the experimental data points. The time constants $\tau_m$ and $\tau_g$ account for the decrease of numbers of ions in the  metastable $2s\,2p\;^3P_0$ level and of ions in the $2s^2\;^1S_0$ ground level, respectively. From the fits to the decay curves of both isotopes, a set of four time constants was obtained (see \citep{Schippers2012} for details) from which the HFI transition rate was derived as
\begin{equation}\label{eq:Ahfi}
    A_\mathrm{HFI} = \frac{1}{\tau_\mathrm{HFI}}
    = \gamma^{(33)}\left[\frac{1}{\tau_m^{(33)}}-\frac{1}{\tau_g^{(33)}}-\frac{n_e^{(33)}}{n_e^{(32)}}\left(
     \frac{1}{\tau_m^{(32)}}-\frac{1}{\tau_g^{(32)}}\right)\right]
\end{equation}
with the superscripts referring to a specific isotope and with the relativistic $\gamma$ factor of the ions and the electron density $n_e$. In Tab.~\ref{tab:comp} the experimental results for $^{47}$Ti$^{18+}$ \cite{Schippers2007a} and $^{33}$S$^{12+}$ \cite{Schippers2012} are compared with the available theoretical HFI $2s\,2p\; ^3P_0 \to 2s^2\; ^1S_0$ transition rates for these ions. The latter and their partly significant deviations from the experimental values will be discussed in the following.

\section{Theoretical $\mathbf{2s\,2p\;^3P_0 \to 2s^2\;^1S_0}$ transition rates}

\subsection{Hyperfine induced transitions}

The hyperfine interaction is the interaction of the atomic electrons with the magnetic and electric fields produced by the atomic nucleus. Consequently, the Hamiltonian \citep{Johnson2011}
\begin{equation}\label{eq:H}
    H_\mathrm{HF} = -\frac{e}{c}\vec{\alpha}\cdot\vec{A}(\vec{r})+e\phi(\vec{r})
\end{equation}
for one electron of charge $e$  contains the electromagnetic potentials $\vec{A}(\vec{r})$ and $\phi(\vec{r})$. These stem dominantly from the nuclear magnetic dipole moment (operator $\vec{\mu}$) and the nuclear electric quadrupole moment (operator components $Q_{ij}$), i.e.,
\begin{equation}
\vec{A}(\vec{r}) = \frac{\vec{\mu}\times\vec{r}}{r^3}\textrm{~~~~~~~and~~~~~~~}\phi(\vec{r}) = \sum_{ij}\frac{r_ir_j}{6r^5}Q_{ij}.
\end{equation}
As described in depth by \citet{Cheng2008a} there are two main issues for the theoretical calculations to deal with: i) the accuracy of the atomic wave functions  and ii) the choice of a perturbative or a nonperturbative method for the calculation of the hyperfine transition matrix. A nonperturbative treatment is required when the level width is of the order of the level-splitting. Since this is not the case for Be-like ions, both methods should yield
the same results for the HFI $2s\,2p\; ^3P_0 \to 2s^2\; ^1S_0$ transition rates.

The first calculations of these rates  were performed  by \citet{Marques1993} for the entire isoelectronic sequence of Be-like ions. They used the multiconfiguration Dirac-Fock (MCDF) method for calculating the atomic wave functions and a nonperturbative complex matrix scheme. The rather large discrepancies between the resulting HFI transitions rates and the experimental values (Tab.~\ref{tab:comp}) are rooted in the neglect of the coupling between the $2s\,2p\;^3P_0$ and $2s\,2p\;^1P_1$ levels (Fig.~\ref{fig:HeBeLevels}) and in an only coarse treatment of electron correlation effects by comparatively moderate configuration expansions. These deficiencies have already been pointed out by \citet{Brage1998a} who obtained different results than \citet{Marques1993} for a number of ions of astrophysical interest (with $Z$ = 6--8, 10--14, 19, 20, 24, 26, and 28), using larger configuration expansions for the wave-functions and a perturbative treatment of the HFI transition matrix which takes the $^3P_0 -^1P_1$ coupling into account.

\citet{Brage1998a} did not calculate $\tau_\mathrm{HFI}$ specifically for $^{47}$Ti$^{18+}$ or $^{33}$S$^{12+}$. However, they found that the nuclear and the electronic parts of the HFI transition rate approximately factorize such that
\begin{equation}\label{eq:AHFIred}
    A_\mathrm{HFI} \approx \mu^2\left(1+\frac{1}{I}\right)\,A_\mathrm{el}
\end{equation}
with the electronic factor $A_\mathrm{el}$ varying only smoothly as function of the nuclear charge (Fig.~\ref{fig:thfi}). Using the fit function of \citet{Brage1998a} for $A_\mathrm{el}$ and the nuclear constants from Tab.~\ref{tab:comp} in Eq.~\ref{eq:AHFIred} yields values for $\tau_\mathrm{HFI}$ which deviate by 73\% and by 18\% from the experimental values for $^{47}$Ti$^{18+}$ and $^{33}$S$^{12+}$, respectively.  A deviation of 18\% could be ascribed to the uncertainty associated with the fit function for $A_\mathrm{el}(Z)$ \citep{Brage1998a}. The reason for the rather large large discrepancy for $^{47}$Ti$^{18+}$ is unclear.

Subsequent to the publication of the experimental HFI $2s\,2p\;^3P_0 \to 2s^2\;^1S_0$ transition rate for $^{47}$Ti$^{18+}$ \cite{Schippers2007a} new theoretical results were published \cite{Cheng2008a,Andersson2009,Li2010}.
\citet{Cheng2008a} performed large scale relativistic configuration interaction (RCI) calculations in order to account accurately for correlation effects. Their results for the entire Be-like isoelectronic sequence are displayed in Fig.~\ref{fig:thfi}.  Moreover, \citet{Cheng2008a} compared the perturbative treatment of the hyperfine transition matrix with the nonperturbative complex matrix and radiation damping methods. They found no significant differences in the results from the perturbative and the nonperturbative radiation damping methods. As already discussed above, this agreement between nonperturbative and perturbative calculations was expected since the relevant level widths are much smaller than the level splittings. The complex matrix approach was shown to systematically yield different results which renders this method unsuitable for the calculation of HFI transition rates in Be-like ions \citep{Cheng2008a}.

The  HFI $2s\,2p\; ^3P_0 \to 2s^2\; ^1S_0$ transition rates calculated by \citet{Andersson2009} for $6\leq Z \leq 22$ and by \citet{Li2010} for $Z$ = 6, 7, 9, 14, 20, 22, 26, 30, 37,45, and 54 generally agree with the results of \citet{Cheng2008a} on the 2\% level, with the exception of a disturbing two-order-of-magnitude difference for $\tau_\mathrm{HFI}$ of $^{103}$Rh$^{41+}$ \cite{Cheng2008a,Li2010}. Within the 5\% experimental uncertainty the results of \citet{Cheng2008a} and \citet{Andersson2009} for $^{33}$S$^{12+}$ agree with the experimental value (Tab.~\ref{tab:comp}). For $^{47}$Ti$^{18+}$ there is a significant $\sim$20\% difference between the new theoretical \citep{Cheng2008a,Andersson2009,Li2010} results and the experimental transition rate.

\subsection{Quenching by  external fields}

In search for the origin of the $\sim$20\% difference (Tab.~\ref{tab:comp}) between the experimental HFI transition rate for $^{47}$Ti$^{18+}$ and the latest theoretical results, which are believed to have converged, \citet{Li2011} investigated in how far the magnetic fields of the storage-ring dipole-magnets (Fig.~\ref{fig:TSR}) could be held responsible for the discrepancy. Such external fields contribute to the electromagnetic potentials in the Hamiltonian (Eq.~\ref{eq:H}) and lead to stronger quenching of the the $2s\,2p\;^3P_0$ level. In addition, the HFI transition rates acquire a dependence on the magnetic quantum number $m_F$ of the excited level. \citet{Li2011} concluded that these effects are too weak at the experimental magnetic flux density of 0.75~T to explain the discrepancy between experimental and theoretical HFI transition rates.

In the rest frame of the ions moving with velocity $\vec{v}$, the $\vec{B}$-field transforms into an electric field $\vec{F} = \vec{v}\times\vec{B}$. Under rather extreme experimental conditions, i.e., for $v=c$ and $B=1.5$~T the electric field strength amounts to $4.5\times10^8$~V~m$^{-1}$. In search for parity-violating effects \citet{Maul1998a} have calculated the rate for the associated quenching of the $2s\,2p\;^3P_0$ level. This rate scales quadratically with field strength $F$. Even for $F=4.5\times10^8$~V~m$^{-1}$ the effect is very weak. The associated lifetimes are of the order of several hours (Fig.~\ref{fig:thfi}).

The effect of external fields on the HFI transition rates has also been investigated experimentally. In the S$^{12+}$ storage-rings experiment \cite{Schippers2012} the ion beam was stored at different magnetic rigidities. The magnetic field strength of the storage ring dipoles was varied by a factor of 2. Within the experimental uncertainties no influence of this $B$-field variation on the measured HFI transition rate was found. At the same time
this shows that the experimental method is robust. It yields the same result under different experimental conditions.

\subsection{E1M1 two-photon transitions}

In Be-like ions the HFI $2s\,2p\;^3P_0 \to 2s^2\;^1S_0$ transition does not compete with any other one-photon transition. However, multi-photon transitions are generally possible. To the best of the author's knowledge there are only two (theoretical) investigations of the E1M1 two-photon $2s\,2p\;^3P_0 \to 2s^2\;^1S_0$ transition. Using nonrelativistic theory \citet{Schmieder1973a} has calculated E1M1 transition rates for Be-like ions with $12\leq Z \leq 20$ (Fig.~\ref{fig:thfi}). Later \citet{Laughlin1980a} --- without referring to the work of \citet{Schmieder1973a} --- derived the following nonrelativistic general expression for the E1M1 transition rate in Be-like ions:
\begin{equation}
    A_\mathrm{E1M1} =\frac{1}{\tau_\mathrm{E1M1}} = A_0 Z^4(E+2\Delta)^2\int_0^{E}d\omega\,\frac{[\omega(E-\omega)]^3}{[\omega(E-\omega)+\Delta(E+\Delta)]^2}\label{eq:AE1M1}
\end{equation}
with $A_0 = 2.867\times10^{-11}$~s$^{-1}$, the $2s^2\;^1S_0\to2s\,2p\;^3P_0$ excitation energy $E$, and the $2s\,2p\;^3P_0\to2s\,2p\;^3P_1$ fine structure splitting $\Delta$. For both $E$ and $\Delta$  absolute values have to be supplied in atomic units. \citet{Laughlin1980a} has not explicitly evaluated the integral but presented an approximate result for $\Delta=0$, i.e.
\begin{equation}
    A_\mathrm{E1M1} = A_0\frac{Z^4}{6}E^5.\label{eq:AE1M1E}
\end{equation}
However, the evaluation of the integral is straightforward and yields \citep{Bernhardt2012a}
\begin{eqnarray}
    A_\mathrm{E1M1} &=& A_0\frac{Z^4}{6}\biggl[E^5-8 E^4 \Delta-68 E^3 \Delta^2-120 E^2 \Delta^3-60  E\Delta^4\nonumber\\
    &&\mathrm{~~~~~~~~~~~~~~}+\frac{12 \Delta^2 \left(3 E^2+10 E \Delta+10 \Delta^2\right) (E+\Delta)^2}{(E+2 \Delta)} \ln\left(\frac{E+\Delta}{\Delta}\right)\biggr].\label{eq:AE1M1D}
\end{eqnarray}
Results from Eq.~\ref{eq:AE1M1E} and \ref{eq:AE1M1D} are plotted in Fig.~\ref{fig:thfi} as function of nuclear charge $Z$. For $Z<14$ both curves agree within 10\%, for higher $Z$ they differ by factors of up to $\sim$2.6. The agreement with the low-$Z$ results of \citet{Schmieder1973a} is within 30\% and 20\%, respectively. All calculated lifetimes $\tau_\mathrm{E1M1}$ are by several orders of magnitude larger than the corresponding HFI induced lifetimes (Fig.~\ref{fig:thfi}). Thus, there is practically no influence of the E1M1 transition on the measured HFI transition rates.

\section{Conclusions and Outlook}

Electron-ion collision experiments at a heavy-ion storage-ring  have provided the first laboratory-experimental HFI $2s\,2p\;^3P_0 \to 2s^2\;^1S_0$ transition rates for Be-like $^{33}$S$^{12+}$ \citep{Schippers2012} and $^{47}$Ti$^{18+}$ \citep{Schippers2007a}. While there is agreement with the results from the latest calculations \citep{Cheng2008a,Andersson2009} for $^{33}$S$^{12+}$, there is a significant 20\% discrepancy between the experimental result for $^{47}$Ti$^{18+}$ and state-of-the-art calculations \citep{Cheng2008a,Andersson2009,Li2010}. The origin of this discrepancy is unclear. As discussed above, there were no significant disturbances of the experimental results by external electromagnetic fields and by the E1M1 two-photon $2s\,2p\;^3P_0 \to 2s^2\;^1S_0$ transition. It seems also unlikely that the tabulated value \cite{Stone2005a} for the magnetic moment of the $^{47}$Ti nucleus
is in error.

In the future, beams of exotic nuclei may become available at the TSR storage ring \cite{Grieser2012}. Measurements of HFI induced transition rates may then be used for determining nuclear magnetic moments, some of which are still known only imprecisely.

The calculation of the E1M1 two-photon rate by \citet{Laughlin1980a} suggests that a corresponding storage-ring measurement may become feasible for high-$Z$ ions with zero nuclear spin where E1M1 lifetimes are shorter than a minute (Fig.~\ref{fig:thfi}). This has already been discussed in more detail by \citet{Bernhardt2012a}. For a thorough planning of such experiments accurate theoretical E1M1 transition rates from state-of-the-art
relativistic methods are required. Preliminary results from relativistic calculations \cite{Fratini} indicate that relativistic effects on the E1M1 two-photon transition rate become drastic at high-$Z$. Moreover, accurate rates for the quenching of the $2s\,2p\;^3P_0$ level by external $B$-fields and by three-photon transitions are needed as well.

\section{Acknowledgments}

The author is indebted to his collaborators {D.~Bernhardt}, {M.~Grieser}, {M.~Hahn}, {C.~Krantz}, {M.~Lestinsky}, {A.~M\"{u}ller}, {O.~Novotn\'y}, D.~A.~Orlov, {R.~Repnow}, {D.~W.~Savin}, {A.~Wolf}, and D.~Yu.  The excellent support by the MPIK accelerator and TSR crews as well as financial support by the Deutsche Forschungsgemeinschaft (DFG, grant no.\ SCHI378/8-1)  is gratefully acknowledged.


\end{document}